\documentclass[sigconf,nonacm]{acmart}
\usepackage{booktabs}
\usepackage{array}
\begin{document}

\title{Era by Eon: Benchmarking Enterprise Agents on Hidden Knowledge}

\author{Benjamin Gruenbaum\qquad Doron Porat\qquad Assaf Natanzon}
\author{Roy Zavida\qquad Chen Dinachi\qquad Or Itzahary}

\renewcommand{\shortauthors}{Gruenbaum et al.}

\begin{abstract}
In the Era by Eon benchmark, each question states the rules for its
answer, and code computes the answer from a generated company's data.
When agents can run code, the four strongest models each answer 22 to
25 of 27 such questions, so the benchmark barely separates
them. We add eight question templates that depend on hidden facts. No
question or document states a hidden fact, and the records that seem
to hold it show something else. Other data implies it. For example,
the sales system says a customer dropped a purchase because of timing.
On a recorded call, the customer blames an outage. For each generated
company, code fills each template and computes an exact answer without
a language model. We evaluate 12 agents. Each pairs a model with an
agent program, which connects it to the company's systems. The best
agent answers 18 of its 24 attempts (three per question) correctly.
Four of the six models answer at most 6 of 24 with any program. The
hardest questions require picking one of several similar records, such
as which of three renewal offers a customer signed. All agents together answered
two such questions correctly in 1 of 84 attempts.
\end{abstract}

\maketitle

\section{Introduction}

Language models can act as agents that answer questions about a
company's business data. An \emph{agent} consists of a language model
and an \emph{agent program}. The agent program gives the model a set
of \emph{tools}, which are functions that call the programming
interfaces (APIs) of the company's systems. The model requests tool
calls, the program runs them and returns the results, and the model
finally composes an answer. Some agent programs also let the model run
code that it writes. One complete attempt to answer one question is a
\emph{run}. Customer and sales data is kept in a customer relationship
management (CRM) system such as Salesforce. Support requests are kept
in a help desk such as Zendesk.

Several benchmarks evaluate such agents against ground truth that comes
from the benchmark's own data. CRMArena generates data for a Salesforce
instance and computes its ground truth from the parameters that its
data generator used~\citep{huang2025crmarena}. CRMArena-Pro extends
it to multi-turn conversations and to tests of
confidentiality~\citep{huang2025crmarenapro}. $\tau$-bench compares
the final state of a database with an annotated goal
state~\citep{yao2024tau}. TheAgentCompany grades long work tasks in a
simulated software company by checking intermediate
steps~\citep{xu2024theagentcompany}. WorkArena and WorkArena++
evaluate browser agents on task templates over the demo data of
ServiceNow~\citep{drouin2024workarena,boisvert2024workarenapp}. The
Era by Eon benchmark generates a fictional company, serves its data
through simulators of business applications, and computes an exact
answer for each question~\citep{gruenbaum2026era}. In the Era by Eon
benchmark, every rule that a question needs is stated in the question
or in a document of the company. We call such questions computable.
BIRD, a benchmark that translates questions into SQL, makes the same
choice explicit: each question comes with an evidence sentence, written
by an expert, that states the outside knowledge that the question
needs~\citep{li2023bird}. Spider~2.0 moves such questions to
enterprise data warehouses~\citep{lei2025spider2}.

This paper continues the Era by Eon benchmark. Its computable questions
now barely separate strong agents. When agents can run code, each of
the four strongest of six models answers between 22 and 25 of 27
computable questions correctly, with one run per question.
Section~\ref{sec:negative} reports these results.

A business question can depend on a fact that no question or document
states. Consider a customer that planned an \emph{expansion}, which is
an additional purchase on top of its existing contract. The CRM tracks
this potential sale as a deal. The customer dropped the expansion, and
the salesperson selected the loss reason ``Timing - revisit next year''
from a fixed list. On a recorded sales call, however, a person at the
customer said that its board had cancelled the expansion because of an
outage of the company's service. The person identified the outage by
the most recent urgent support ticket that the customer had opened
before the deal closed. Urgent is the highest priority of a support
ticket. An agent that reads only the CRM answers that the cause was
timing, and this answer is wrong. To answer correctly, the agent must
find the remark among many calls and recognize that it states the
cause. It must then identify the ticket in the help desk. We call a
fact of this kind \emph{hidden knowledge}. This paper adds eight
question templates to the Era by Eon benchmark. A template fixes the
wording of a question and the rule that gives its answer. For a
generated company, the benchmark selects a customer, a deal, or a
salesperson from the company's data. It fills the names of this
entity into the template and computes the answer. A company that is
generated with a different random seed has different data. The
benchmark therefore selects different entities and computes different
answers.

We evaluated 12 agents on the eight questions of one company, with
three runs per question. Each agent pairs one of six language models
with one of two agent programs, and every model ran in both programs. The best agent answered 18 of its 24 runs
correctly. Claude Fable 5.1 and GPT-6 Astra are the two models with
the most correct runs. For these two models, the ability to run code
did not raise the number of correct runs. The hardest questions ask
the agent to pick one record among several that look alike. For
example, a customer received three renewal offers and signed one of
them. Each offer states the last day on which its price holds, and
only one offer was still valid on the day that the customer signed.
Two questions of this kind were answered correctly in 1 of the 84 runs
that all agents made on them.

This paper makes three contributions:
\begin{enumerate}
\item We define hidden knowledge, distinguish three kinds of it, and
  state four conditions under which a question that depends on it has
  one exact answer (Section~\ref{sec:hidden}).
\item We add eight question templates to the Era by Eon benchmark. We
  describe how the benchmark generates a question and its answer from
  each template without a language model, and how the question changes
  with the seed (Sections~\ref{sec:build} and~\ref{sec:questions}).
\item We evaluate 12 agents on the eight questions of one company and
  report where they fail (Sections~\ref{sec:setup}
  and~\ref{sec:results}).
\end{enumerate}

\section{Background: the Era by Eon benchmark}
\label{sec:background}

Era by Eon is a benchmark developed at Eon, the company of the authors.
It generates a fictional company and simulates the business
applications that the company uses~\citep{gruenbaum2026era,gruenbaum2026eragen}.
The input is a specification with five parts. They are an industry, a
size class from micro to hyperscale, a business model such as selling
to other businesses, a list of business applications, and a random
seed. Generation is deterministic, so the same specification always
produces the same company. The generator first builds a single data
structure, the \emph{entity graph}. The graph holds the company's
employees, customers, deals, support tickets, recorded calls, chat
messages, and documents.

Each business application is served by a simulator. A simulator
implements two interfaces of the real product. The first is its REST
API, which is the web interface through which programs use the
product. The second is its Model Context Protocol (MCP) server, which
is a standard interface through which a language model calls tools.
Salesforce and HubSpot are CRM systems. Zendesk is a help desk, and
Jira tracks engineering work. Gong records sales calls and stores their
transcripts. Slack is a chat system, and Amazon S3 serves as the file
store. We call each item that a system stores a \emph{record}. For
example, a deal in Salesforce, a ticket in Zendesk, and a call with its
transcript in Gong are records. We say that data is \emph{served} when
a simulator returns it to an agent. Agents have read-only access.

All simulators read from the same entity graph. As a result, a
customer that appears in Salesforce also appears in Zendesk and in
Gong with consistent details. Together, the simulators form a
\emph{digital twin} of the company. A digital twin is a software
replica that behaves like a set of real systems. Here, each simulator
behaves like its real product, and the data describes the fictional
company. The data of the twin comes from one consistent entity graph.
Its interfaces are the REST APIs and MCP servers of the real products.
In a few places, the generator makes two systems disagree on purpose,
in order to build questions of this paper.

Code computes the correct answer to each question from the served
data. We call this answer the \emph{key}. The \emph{grader} is the
program that compares an agent's answer with the key. The generated
companies are available at
\url{https://console.era.eon.io}~\citep{eraconsole}.

We use the following business terms. An \emph{account} is a customer
company, and a \emph{contact} is a person who works at an account. A
\emph{deal} is a potential sale with an amount and an expected close
date. A deal moves through \emph{stages}, such as Qualification,
Proposal, and Negotiation, until it is won or lost on its close date.
A \emph{contract} records the terms that an account has signed, and
it covers a period called its \emph{term}. A \emph{renewal} extends a
contract. A \emph{proposal} is a document that states the price and
terms of a renewal. The \emph{pipeline} is the set of open deals. The
\emph{committed pipeline} is the amount that the sales team expects to
close from the pipeline, and each company computes it with its own
method. A \emph{ticket} is a support request. Jira stores work items
called \emph{issues} and groups them into \emph{projects}. Each issue
has a priority, which in our data is Highest, High, Medium, or Low.
The \emph{roadmap} is the planned engineering work in Jira. All
amounts are in US dollars.

Unless stated otherwise, the experiments use one generated company. It
is a large financial-technology (fintech) company that sells to other
businesses, generated with a fixed seed. The generator's size scale
labels this company as having 1{,}000 employees. The generated data
contains records for 40 of them, who are the users of the business
systems. Of these 40, 17 work in engineering, 13 in customer support,
8 in sales, and 2 in presales. The company has 250 accounts, 4{,}190
contacts, 198 contracts, 2{,}500 deals in Salesforce, 3{,}000 tickets,
3{,}506 recorded calls, 9{,}880 Jira issues, 1{,}918 Slack messages,
and 2{,}367 documents in the file store. HubSpot holds copies of
2{,}464 of the Salesforce deals. Some copies differ from the Salesforce
deal in amount or close date.

\section{Limits of computable questions}
\label{sec:negative}

This paper continues the Era by Eon benchmark
paper~\citep{gruenbaum2026era}, which we call the benchmark paper. The
benchmark paper evaluated agents on computable questions. We call a
question \emph{computable} when every rule that is needed to compute
its answer is stated in the question or in a document in the company's
file store. A \emph{rule} fixes a choice that the answer depends on.
For example, a rule can fix the time window over which tickets are
counted. This section first summarizes the benchmark paper. It then
reports how agents that can run code perform on the current computable
questions. These results motivate the questions of this paper.

The benchmark paper used its own agent program, which we call the
\emph{Era agent program}. In that paper, the model starts with three
tools. The first lists the systems of the company. The second
describes the tools of a chosen system's MCP server, and the third
calls one of these tools. The prompt does not list the tools of the
MCP servers, so the model must discover which of them answer the
question. The Era agent program cannot run code. The model must
therefore read the records that it needs in tool results and compute
the answer itself.

A run consists of \emph{turns}. In each turn, the model receives the
question, its own earlier responses, and the results of all earlier
tool calls. It then writes one response. The response either requests
tool calls or gives the final answer. When the response requests tool
calls, the agent program runs them, and the next turn begins with their
results. For example, in one turn the model may request the list of
accounts from Salesforce. In the next turn, it may read this list and
request the tickets of one account from Zendesk. One response can
request several tool calls, so an agent program limits turns and tool
calls separately. In the benchmark paper, a run was limited to 25
turns, 40 tool calls, and 8{,}000 characters per tool result.

In the benchmark paper, nine models answered 33 computable questions
about a small fintech company. The company had 40 accounts and 240
support tickets. Each model answered each question three times. The
share of correct runs ranged from 42.4\% for GPT-5.4 mini to 76.8\% for
Claude Opus 4.8. The benchmark paper tested each of the 36 pairs of
models for a statistically significant difference, with a correction
for multiple comparisons. Only three pairs showed one, and each of them
compared GPT-5.4 mini with a stronger model. The benchmark paper also
labeled each question with the capabilities that it tests. Averaged
over the nine models, the share of correct runs was 3.7\% on questions
that chain several lookups and 92.6\% on questions that only select
records by a condition. In a chain of lookups, each lookup uses the
result of the previous one. Aggregating every record of one type and
ranking records were also among the hardest capabilities.

A script can perform these operations directly. It can follow a chain
of lookups, read every record of one type, and sort the results. Only
its printed output goes back to the model, so the model does not have
to read thousands of records itself. We therefore asked whether agents
that can run code answer computable questions correctly. For this
purpose, this paper uses two agent programs. The first is a new
version of the Era agent program, and the second can run code. We
describe them in turn.

The new version of the Era agent program differs from the version of
the benchmark paper in two ways. First, the agent can call the REST
API of each simulator as well as its MCP server. The version of the
benchmark paper used only the MCP servers. Second, a run may be
longer. It may use up to 165 turns and 165 tool calls, and a tool
result may have up to 32{,}000 characters. Like the earlier version,
the new version cannot run code. In the rest of this paper, the Era
agent program means this new version.

The second agent program is built with LangGraph~\citep{langgraph}, an
open-source library for building agents that call tools. We call it
the \emph{LangGraph program}. It has the same tools as the Era agent
program and one more tool, which runs Python code in a \emph{sandbox}. A
sandbox is an isolated Python interpreter. The sandbox has no network
access of its own. A script sends its requests to the agent program,
which forwards them to the REST APIs of the simulators. A script can
therefore page through thousands of records and compute an answer. The
two programs thus reach the same interfaces of the simulators, and
they differ in code execution. A turn has the same meaning in both
programs. In the LangGraph program, a response can also ask the sandbox
to run a script, and the script's printed output returns as the result
of that tool call. The LangGraph program stops a run when the model
gives its answer or after 150 turns. This limit is close to
the 165 turns of the Era agent program.

We use 27 computable questions of the current version of the Era by
Eon benchmark. These questions differ from the 33 questions of the
benchmark paper in two ways. First, they were written for the large
company of Section~\ref{sec:background}. Second, each of the 33
questions asks for one fact. Most of the 27 questions ask for several
values about one subject, and a run is correct only when every value
is correct. The following is one of the 27 questions: ``For each account tier, what
percentage of that tier's accounts have at least one urgent support
ticket that is still open?'' A \emph{tier} is one of three size classes
of accounts. From largest to smallest, the tiers are Enterprise,
Mid-Market, and SMB, which stands for small and medium-sized
businesses. To answer, the agent must join the 250 accounts in
Salesforce with the 3{,}000 tickets in Zendesk. It must then give three
percentages, one for each tier.

We ran six models from three providers on these questions. They are
Claude Fable~5.1 and Claude Sonnet~5 from Anthropic; GPT-6 Astra,
GPT-5.6 Sol, and GPT-5.6 Luna from OpenAI; and DeepSeek-V3.2 from
DeepSeek. We evaluate the same six models on the questions of this
paper. On the computable questions, each model made one run per
question in each program that it ran in. Table~\ref{tab:computable}
shows which programs these were.

Table~\ref{tab:computable} shows the number of correct answers. With
the LangGraph program, the four strongest models answered between 22
and 25 of the 27 questions correctly. Two of these four models also ran
in the Era agent program, which cannot run code. There, GPT-6 Astra
answered 8 questions and GPT-5.6 Sol answered 1. For these two models,
code execution added 16 and 21 correct answers. GPT-5.6 Luna answered 8
questions with code and none without. DeepSeek-V3.2 answered one
question without code and none with it. With code, the four strongest
models are within three questions of each other. With one run per
question, these counts give little evidence for ranking the four
models. The scores cannot be compared with those of the benchmark
paper, because the company, the questions, and the limits of the Era
agent program differ.

\begin{table}[!h]
\caption{Number of correct answers on the 27 computable questions of
the current benchmark, with one run per question. The Claude models
were not run in the Era agent program on these questions, to limit
cost.}
\label{tab:computable}
\begin{tabular}{lcc}
\toprule
Model & Era agent program & LangGraph program \\
 & (no code) & (with sandbox) \\
\midrule
Claude Fable 5.1 & not run & 25 \\
Claude Sonnet 5 & not run & 25 \\
GPT-6 Astra & 8 & 24 \\
GPT-5.6 Sol & 1 & 22 \\
GPT-5.6 Luna & 0 & 8 \\
DeepSeek-V3.2 & 1 & 0 \\
\bottomrule
\end{tabular}
\end{table}

A script can read every record that a computable question needs and
apply the rules that the question states. This likely explains why
code execution added 16 and 21 correct answers for GPT-6 Astra and
GPT-5.6 Sol. The questions of this paper depend on a fact that no
question or document states. The records that seem to hold the answer
show something else, and other served data implies the fact. An
example is a remark by a customer on a recorded call.
Section~\ref{sec:hidden} defines such facts.

\section{Hidden knowledge}
\label{sec:hidden}

Section~\ref{sec:negative} showed that the strongest agents answer most
computable questions correctly when they can run code. In a computable
question, the question or a document states every rule that the answer
needs. This paper adds questions whose answer depends on a fact that
neither the question nor any document states. We call such a fact a
\emph{hidden fact}. This section first explains hidden facts with one
example. It then describes three kinds of hidden facts. Finally, it
states four conditions that every question of this paper meets, so that
each question has one exact answer.

Our example is a question about a service credit. A service credit is
a discount that a customer receives when the company does not provide
the service that the contract promises. The benchmark generates this
question from a template, which Section~\ref{sec:build} describes. For
the company of Section~\ref{sec:background}, the template selects the
customer Dovewood Analytics. For a company that is generated with
another seed, the template selects another customer. The agent
receives the following question:
\begin{quote}
\raggedright
What SLA credit is Dovewood Analytics actually owed on its current
contract? Answer a JSON object with exactly these keys:
\texttt{credit\_pct} (the credit as a bare percentage number),
\texttt{effective\_month} (the first month it applies,
\texttt{YYYY-MM}), and \texttt{months\_covered} (how many calendar
months from that month through the month the served contract ends,
inclusive).
\end{quote}

The question asks about a contract, and Salesforce stores the
company's contracts. Dovewood Analytics has one contract, whose term is
19 September 2023 to 19 September 2024. The contract record has a field
that holds the service credit in percent. For this contract, the field
holds 0. An agent that relies on this field answers that Dovewood
Analytics is owed no credit. This answer is wrong, because the credit
was promised on a recorded call and Salesforce does not record the
promise.

The promise is in Gong. On 22 August 2024, Ana Dalewick, the
salesperson who manages the Dovewood Analytics account, took part in a
call with two employees of Dovewood Analytics. At the end of the call,
she says:
\begin{quote}
On the credit, here is what I'm putting on the table, and I'll own
that it is not in the contract paperwork yet: five points off for
every calendar month of this term in which you've had to open an
urgent-priority ticket with our desk, capped at fifteen, and it runs
from the first month that happened. Hold me to it.
\end{quote}
The remark states a rule but no dates. The agent must apply the rule
to the tickets in Zendesk. Zendesk stores eleven tickets of Dovewood
Analytics, and four of them have the priority urgent. They were opened
on 6 December 2023, 1 March 2024, 31 May 2024, and 18 September 2024.
They fall in four different months of the contract term. At five
points per month, the credit would be 20 percent, so the cap of 15
percent applies. The credit starts in December 2023, the month of the
first urgent ticket. From December 2023 through September 2024, the
month in which the contract ends, there are 10 months. The key is
therefore a credit of 15 percent that starts in December 2023 and
covers 10 months.

\subsection{Definition}

The promised credit rule is an example of a hidden fact. A hidden fact
has three properties. First, the key depends on it. Second, neither the
question nor any document in the file store states it, and the records
that the question asks about do not show it. Third, other served data
gives evidence for it. In the example, the key depends on the promised
rule. Neither the question nor any document mentions the promise, and
the contract record shows a credit of 0. The evidence is the remark on
the call in Gong. The second property separates hidden facts from the
rules of computable questions. The question or a document states those
rules.

\subsection{Three kinds of hidden knowledge}

The eight questions of this paper fall into three kinds. In four
questions, the evidence for the hidden fact is a remark on one recorded
call. The service-credit question is one of them. In three questions,
several records seem to answer the question, and the agent must find
out which of them the company relies on. In one question, the agent
must work out from an old report how the company computes an amount.
Section~\ref{sec:questions} describes each question.

\subsection{Conditions for an exact answer}

A question that depends on a hidden fact must still have one answer
that a program can grade. The benchmark checks four conditions for
every question. Section~\ref{sec:build} describes how.

\begin{enumerate}
\item \textbf{The evidence is served.} The benchmark computes the key
  from the data that the simulators return. In the example, this data
  is the remark in Gong, the tickets in Zendesk, and the contract in
  Salesforce.
\item \textbf{The answer is unique.} Under the rule that the key uses,
  the data allows exactly one answer. In the example, Dovewood
  Analytics must have exactly one contract, so that the question names
  one contract.
\item \textbf{The obvious answer is wrong.} An agent that reads only
  the record that the question asks about gives a wrong answer. In the
  example, this record is the contract, which shows a credit of 0. The
  key is a credit of 15 percent.
\item \textbf{The answer is exact.} The answer consists of two or three
  values, such as a number, a month, or a name. The question states the
  form of each value. For example, the service-credit question asks for
  the month in the form 2023-12. An answer is correct only if all of
  its values are equal to the key.
\end{enumerate}

\section{From templates to questions}
\label{sec:build}

Section~\ref{sec:hidden} showed one question and traced its answer
through three systems. Nobody wrote that question by hand. It came
from a \emph{template}, which is a recipe that produces one question of
its kind for any generated company. This paper adds eight templates to
the benchmark, and each of them produces one question for a company.
This section tells how a template becomes a question. We follow the
service-credit question of Section~\ref{sec:hidden} from the moment
the company is generated to the moment the agent reads the question.
At the end of the section, we generate a second company with another
seed and show that the same template produces a different question
with a different answer. Figure~\ref{fig:template} shows the
service-credit template, part by part.

\begin{figure*}[tb]
\small
\begin{tabular}{@{}>{\bfseries}p{0.15\textwidth}p{0.82\textwidth}@{}}
\toprule
\multicolumn{2}{@{}l}{\textbf{Template: service credit}\hfill Systems: Salesforce, Zendesk, Gong}\\
\midrule
Search rule & a contract whose customer opened urgent tickets in at least
two different months of the contract term; take the first candidate in a
fixed order\\
\addlinespace
Hidden fact & \texttt{credit\_pct} = 5 points per such month, capped at 15\newline
\texttt{effective\_month} = the first such month\newline
written to the answer file only; no system changes\\
\addlinespace
What the systems show & Salesforce, object Contract, field
\texttt{SLA\_Credit\_Pct\_\_c} = 0\\
\addlinespace
Clue & call: the customer's last recorded call after the first urgent month
and before the contract ends\newline
speaker: an employee of the company who is on that call\newline
text: ``On the credit, here is what I'm putting on the table, and I'll own
that it is not in the contract paperwork yet: five points off for every
calendar month of this term in which you've had to open an urgent-priority
ticket with our desk, capped at fifteen, and it runs from the first month
that happened. Hold me to it.''\\
\addlinespace
Question & ``What SLA credit is \{customer\} actually owed on its current
contract? Answer a JSON object with exactly these keys: \texttt{credit\_pct},
\texttt{effective\_month}, \texttt{months\_covered}.''\\
\addlinespace
Checks & the contract's credit field is 0\newline
the customer has exactly one contract\newline
at least two months of the term have an urgent ticket\newline
recomputing the hidden fact from the records gives the same values\\
\addlinespace
Key & \texttt{credit\_pct}, \texttt{effective\_month},
\texttt{months\_covered}; the answer is correct only if all three match\\
\bottomrule
\end{tabular}
\caption{The service-credit template. The first four parts run in the
generator and shape the company. The last three run when the benchmark
is assembled and produce the question and its key.}
\label{fig:template}
\end{figure*}

Everything begins with the company. The generator builds it from the
specification and the seed, as Section~\ref{sec:background} described.
It creates the employees, the customers, their contracts, their
support tickets, the recorded calls with their transcripts, and the
documents in the file store. No template has run at this point.
Dovewood Analytics is one customer among 250. It has a one-year
contract, and it has opened eleven support tickets, four of them
urgent. Its contract shows a service credit of zero, and none of its
23 recorded calls mentions a credit. Nothing in the company yet points
to the question that the agent will later receive.

The service-credit template now searches the finished company for a
customer to ask about. It looks for a customer whose support tickets
show a pattern that would justify a credit. The pattern is urgent
tickets in at least two different months of the same contract term.
Many customers of this company fit, and the template needs one of
them. It sorts the candidates in an order that depends only on their
identifiers and takes the first. This order is the same on every run,
so the same company always produces the same question. In this
company, the template arrives at the contract of Dovewood Analytics.
That contract ran from 19 September 2023 to 19 September 2024, and the
four urgent tickets of Dovewood Analytics fall in four different
months of that term.

With the customer chosen, the template invents what the company
promised. It counts the months of the term in which Dovewood
Analytics opened an urgent ticket. Four months qualify, which gives
five points each and twenty in total, and the promise caps the credit
at fifteen percent. The credit runs from December 2023, the first of
those months. The template writes these two values and the rule that
produced them into a file that belongs to the benchmark. This file
never reaches any simulator, so no agent can read it. It holds the
answer that the grader will later expect. The company's own systems
are left as they are, and the contract of Dovewood Analytics still
shows a service credit of zero.

The promise now exists only in the benchmark's file, so the template
has to give the agent a way to discover it. It does so with one
sentence spoken on a recorded call. The template goes through the
calls that Dovewood Analytics held between December 2023, the first
month with an urgent ticket, and the end of the contract. There are
eight such calls. It takes the last of them, the call of 22 August
2024 in which Ana Dalewick discussed a new deal with two employees of
Dovewood Analytics. To the end of that transcript, the template adds
the remark that Section~\ref{sec:hidden} quoted, spoken by Ana
Dalewick. Nothing else in the call changes, and the earlier turns do
not lead up to the remark. This is the only change that any template
makes to the company. The remark states the rule of the promise but
none of its numbers. The agent has to work out the months, the
percentage, and the duration from the tickets and the contract.

The company is now complete, and the simulators take over. Each of
them serves its part of the company to the agent: Salesforce serves
the contract with its credit field of zero, Zendesk serves the eleven
tickets, and Gong serves the 23 calls, including the one that now ends
with the remark. Before the agent sees anything, the benchmark
performs its own check. It fills the name Dovewood Analytics into the
question text and computes the answer again, this time only from what
the simulators serve. It requires that Dovewood Analytics has exactly
one contract, that the credit field of this contract holds zero, and
that the recomputed credit and starting month agree with the values in
its file. If any of these checks fails, the benchmark drops the
question for this company rather than ask a question with a doubtful
answer. If all of them pass, the agent receives the question, and the
grader later compares the agent's three values with fifteen percent,
December 2023, and ten months.

Two things hold for every step of this story. First, no language
model takes part in it. Code generates the company, including every
transcript and document, code chooses the customer, code writes the
remark, and code computes the answer. The questions and their answers
therefore contain no model errors. Second, every step is
deterministic, so the same specification and seed always give the
same company, the same customer, and the same answer. The other seven
templates follow the same path. Each of them needs at least two
customers of the right kind to choose from. We measured these counts
for the fintech companies only, and the four companies of the sizes
mid, large, enterprise, and hyperscale have enough for every template.
The benchmark therefore adds the eight questions to these four
companies.

Two things hold for every step of this story. First, no language
model takes part in it. Code generates the company, including every
transcript and document, code chooses the customer, code writes the
remark, and code computes the answer. The questions and their answers
therefore contain no model errors. Second, every step is
deterministic, so the same specification and seed always give the
same company, the same customer, and the same answer. The other seven
templates follow the same path. Each of them needs enough customers of
the right kind to choose from. Companies of the size mid or larger
have enough data for all eight questions, and smaller companies have
too few customers of some kinds.

To see what the seed changes, we generated the same company a second
time, with seed 43 instead of 42 and everything else unchanged. The
second company also has 250 customers and 2{,}500 deals, but most
customers have other names, and their contracts, tickets, and calls
differ. The service-credit template ran on this company as described
above and chose the customer Swinburn Holdings. Its contract, like
Dovewood's, has urgent tickets in at least two months of its term, and
the template's fixed order placed it first. The question text is the
same apart from the name, and the remark is the same sentence on a
different call. The answer depends on the customer. A customer with
urgent tickets in exactly two months would be owed ten percent. Both
Dovewood Analytics and Swinburn Holdings had more such months, so both
reach the cap of fifteen percent, from December 2023 in both cases.
The number of months differs: ten for Dovewood Analytics and thirteen
for Swinburn Holdings, because the contract of Swinburn Holdings ends
three months later.

\section{The seven other templates}
\label{sec:questions}

Figure~\ref{fig:template} showed one template. The other seven follow
the same structure and fall into the three kinds of
Section~\ref{sec:hidden}. This section explains each kind and then
states, for each template of that kind, what it looks for, what the
truth is, and what the question asks. The names in bold are the names
by which the rest of the paper refers to the questions.

\subsection{A remark on a call}

In four templates, the clue is a remark on a recorded call. The
customer's records are complete and consistent, and they suggest an
answer. Salesforce shows a loss reason for a lost deal, a recent
activity date for a contact, or a credit of zero on a contract. The
template leaves these records as they are. It adds one remark to one
of the customer's recorded calls. The remark states a rule for
computing the real answer from the records. For the service credit,
the rule is five points per month with an urgent ticket, capped at
fifteen. The answer itself is never said on the call and is never
stored in any system. It is stored only in the answer file. The agent
has to find the remark, understand the rule, and compute the answer
from the records. Figure~\ref{fig:template} showed one of the four
templates. The other three follow.

\textbf{Lost deal.} The customer says on a call that a lost deal was
waiting on a roadmap story of the engineer who handled most of its
bugs, and the question asks which story. For the company of Section~\ref{sec:background}, the agent
receives:
\begin{quote}\raggedright
The Soleris Partners opportunity \texttt{Soleris Partners --- Platform Rollout} (closed 2024-07-31) was lost. Answer a JSON object with exactly these keys: \texttt{blocker} (the key of the roadmap issue in the tracker, e.g. \texttt{PLAT-12}, that the deal was actually waiting on when it was lost) and \texttt{amount} (the opportunity's amount as the CRM serves it, a bare integer).
\end{quote}

\textbf{Departed contact.} The customer says on a call that a regular
participant has left the month after their last call, and the question
asks who left and when. For the company of Section~\ref{sec:background}, the agent
receives:
\begin{quote}\raggedright
Which of the Turquoise Systems contacts who used to be on the calls is no longer at the company? Answer a JSON object with exactly these keys: \texttt{contact} (their full name as the CRM spells it), \texttt{departed} (the month they left, \texttt{YYYY-MM}), and \texttt{still\_active\_in\_crm} (true if the CRM still shows activity for them dated 2024-08 or later).
\end{quote}

\textbf{Dropped expansion.} The customer says on a call that it dropped
an expansion because of the outage behind its last urgent ticket, and
the question asks for the cause, the ticket, and the lost amount. For the company of Section~\ref{sec:background}, the agent
receives:
\begin{quote}\raggedright
Seraphis Group passed on an expansion: the opportunity \texttt{Seraphis Group --- Data Retention Add-on} closed lost and the account renewed on its existing terms instead. Answer a JSON object with exactly these keys: \texttt{cause} (one of \texttt{requirements}, \texttt{pricing}, \texttt{incident}, \texttt{competitor}, \texttt{budget} --- what actually drove the decision), \texttt{ticket} (the support ticket's external id, e.g. \texttt{fin-large-ticket-1-0}, that the cause traces to, or null), and \texttt{arr\_delta} (the lost opportunity's CRM amount minus the annual value the renewal proposal in the document store states for the served contract's term, a bare integer, negative when the renewal is larger).
\end{quote}

\subsection{Several records, one is right}

In three templates, no call carries a clue. The company holds two or
more records that all seem to answer the question, and only one of
them is the record that the company relies on. The template writes
which one into the answer file and changes nothing in the systems. The
evidence is already in the records: a property that names the system
of record, the date on which a project stopped changing, or a line on
a document that says until when its price holds. The agent has to
notice this evidence and set the other records aside.

\textbf{Conflicting amounts.} Salesforce and HubSpot record one deal
with different amounts, and the question asks which amount the
forecast commits and for which fiscal quarter. For the company of Section~\ref{sec:background}, the agent
receives:
\begin{quote}\raggedright
The Berylpeak Labs deal \texttt{Berylpeak Labs --- Initial Rollout} created on 2023-05-09 is tracked in both CRMs with different amounts and close dates. Which number is the one committed in the forecast, and for when? Answer a JSON object with exactly these keys: \texttt{amount} (the committed amount, a bare integer) and \texttt{quarter} (the fiscal quarter it was committed for, in the form \texttt{FY2025Q2}).
\end{quote}

\textbf{Roadmap.} Jira holds an old roadmap project that stopped
changing and a new one that is live, and the question asks which
engineer has the most open roadmap stories. For the company of Section~\ref{sec:background}, the agent
receives:
\begin{quote}\raggedright
Which engineer currently carries the most open roadmap stories, and how many? Answer a JSON object with exactly these keys: \texttt{engineer} (their display name as the tracker serves it) and \texttt{open\_stories} (the number of Story issues assigned to them that are not Done, a bare integer).
\end{quote}

\textbf{Signed proposal.} The file store holds several rounds of one
renewal proposal with different uplifts, and the question asks which
round the customer signed and what uplift it carries. For the company of Section~\ref{sec:background}, the agent
receives:
\begin{quote}\raggedright
Which renewal proposal round did Evidora Logistics actually execute for its current contract, and what uplift did it agree to? Answer a JSON object with exactly these keys: \texttt{term\_start} (the term start date that proposal states, YYYY-MM-DD) and \texttt{uplift\_pct} (the uplift percentage its commercial summary states, a bare integer).
\end{quote}

\subsection{A method in an old report}

In one template, the clue is a report in the file store. The report
shows the results of a computation on the day it was written, but it
does not state the rules of the computation. The company's people know
the rules and apply them to the current records. The template writes
the rules and today's result into the answer file and changes nothing
in the systems. The agent has to recover the rules from the rows of
the report and run them again on the records as they are today.

\textbf{Pipeline.} A pipeline review lists a salesperson's deals with
a committed amount for each, and the question asks what the
salesperson's committed pipeline is today under the review's own
rules. For the company of Section~\ref{sec:background}, the agent
receives:
\begin{quote}\raggedright
RevOps' pipeline review for Sara Novak in the document store is a snapshot. What is Sara Novak's committed pipeline today under that review's own commit model, run against Salesforce as it stands now? Answer a JSON object with exactly these keys: \texttt{committed\_pipeline} (the total the model commits, a bare integer in dollars) and \texttt{opportunities\_counted} (how many of their opportunities the model counts, the way the review's total row counts them, a bare integer).
\end{quote}

\section{Experimental setup}
\label{sec:setup}

We now run agents on the eight questions of Section~\ref{sec:questions}.
The question we want to answer is simple: can an agent that reads the
company's systems find the hidden facts, and does the ability to run
code help? This section states what we ran and how we scored it.
Section~\ref{sec:results} gives the results.

\textbf{Agents.} An agent is one language model paired with one agent
program. We use the six models of Section~\ref{sec:negative} and the
two agent programs of Section~\ref{sec:negative}, the Era agent
program, which cannot run code, and the LangGraph program, which can.
Every model ran in both programs. This gives twelve agents.

\textbf{Limits.} The two programs keep the limits of
Section~\ref{sec:negative}. Two further limits apply in these
experiments. The Era agent program stops a run after 150{,}000 tokens
of model output. The LangGraph program stops a run after one hour, and
one script in its sandbox may run for at most five minutes.

\textbf{Access.} The agent can read from the systems and cannot write
to them. The sandbox cannot see the benchmark's own files: the
generated data, the keys, and the answer file. The agent reaches the
company only through the systems, like an employee.

\textbf{Runs.} Each agent answers each question three times. A run ends
when the agent gives its answer or reaches a limit. Each agent
therefore has 24 runs.

\textbf{Grading.} The agent gives its answer as a JSON object with the
fields that the question names. A run is correct when every field
equals the key. A run with some correct fields is incorrect. An
abstention is incorrect, because every question has an answer. A run
that reaches a limit without an answer is incorrect.

\textbf{What we report.} For each agent, we report the number of
correct runs out of 24 and a 95\% confidence interval for the share of
correct runs, computed with the Wilson method. We also report how many
runs were abstentions. For comparison, we give the number of the 27
computable questions of Section~\ref{sec:negative} that the agent
answered correctly, one run per question. This shows whether an agent
that is strong on computable questions is also strong on hidden
knowledge.

\section{Results}
\label{sec:results}

Table~\ref{tab:results} gives the results. Two models stand apart.
Claude Fable 5.1 answered 18 of its 24 runs correctly in the Era agent
program and 15 in the LangGraph program. GPT-6 Astra answered 13 and
12. The other four models answered at most 6 of 24 in either program.
We now look at the three kinds of question in turn, then at what the
agents did on the hardest ones.

\begin{table*}[tb]
\caption{Results of the 12 agents on the eight questions, with three
runs per question. The interval is a 95\% Wilson interval for the share
of correct runs, in percent. Abstained counts runs out of 24. The last
column counts the computable questions that the agent answered
correctly, out of 27, with one run per question. ``Not run'' means that
the agent did not run on the computable questions.}
\label{tab:results}
\small
\begin{tabular}{llccccc}
\toprule
Model & Agent program & Correct runs (of 24) & 95\% interval & Abstained & Computable (of 27) \\
\midrule
Claude Fable 5.1 & Era           & 18 & 55.1--88.0 & 0  & not run \\
Claude Fable 5.1 & LangGraph     & 15 & 42.7--78.8 & 0  & 25 \\
GPT-6 Astra      & Era           & 13 & 35.1--72.1 & 5  & 8 \\
GPT-6 Astra      & LangGraph     & 12 & 31.4--68.6 & 5  & 24 \\
Claude Sonnet 5  & Era           & 6  & 12.0--44.9 & 2  & not run \\
Claude Sonnet 5  & LangGraph     & 6  & 12.0--44.9 & 0  & 25 \\
GPT-5.6 Sol      & LangGraph     & 2  & 2.3--25.8  & 0  & 22 \\
GPT-5.6 Sol      & Era           & 1  & 0.7--20.2  & 1  & 1 \\
GPT-5.6 Luna     & Era           & 1  & 0.7--20.2  & 7  & 0 \\
GPT-5.6 Luna     & LangGraph     & 0  & 0.0--13.8  & 7  & 8 \\
DeepSeek-V3.2    & Era           & 0  & 0.0--13.8  & 14 & 1 \\
DeepSeek-V3.2    & LangGraph     & 0  & 0.0--13.8  & 3  & 0 \\
\bottomrule
\end{tabular}
\end{table*}

\textbf{A remark on a call.} On the four questions of this kind, the
twelve agents answered 54 of 144 runs correctly. Claude Fable 5.1 in
the Era agent program answered all twelve of its runs. In the LangGraph
program it answered nine, and each GPT-6 Astra agent also answered
nine. The departed-contact question was the easiest of the eight: 21 of
its 36 runs were correct. The lost-deal, dropped-expansion and
service-credit questions were correct in 10, 13 and 10 of 36 runs.

\textbf{Several records, one is right.} These three questions were the
hardest. The conflicting-amounts question was correct in 7 of 36 runs:
all six runs of Claude Fable 5.1 and one run of GPT-6 Astra in the Era
agent program. Here the evidence is a property on the HubSpot record
that names Salesforce as the system of record. The roadmap question
was correct in none of its 36 runs. The signed-proposal question was
correct once, by GPT-5.6 Luna in the Era agent program. On these two
questions, ten of the twelve agents had both values wrong in every
run. The six runs that Claude Fable 5.1 failed in the Era agent
program are exactly its runs on these two questions.

\textbf{A method in an old report.} The pipeline question was correct
in 12 of 36 runs. All twelve came from the four agents of Claude Fable
5.1 and GPT-6 Astra, and each of them answered all three of its runs.

\textbf{How much the agents searched.} For each agent and question, we
averaged the number of tool calls over the three runs. A call to the
sandbox may read many records, so this counts actions, not data. For
the four agents of Claude Fable 5.1 and GPT-6 Astra, the roadmap and
signed-proposal questions had the two lowest averages of the eight
questions, between 7 and 17 calls. On the other six questions, the
same agents made between 17 and 109 calls. None of the four abstained
on the two hardest questions. They answered quickly, and wrongly.

\textbf{Does code help?} On the computable questions of
Section~\ref{sec:negative}, code helped GPT-6 Astra a great deal: 24
correct in the LangGraph program against 8 in the Era agent program.
On the eight questions of this paper, it did not: 12 against 13.
Claude Fable 5.1 answered 15 runs with code and 18 without.

\textbf{The weaker models.} Claude Sonnet 5 answered 6 runs in each
program. DeepSeek-V3.2 answered none. In the Era agent program it
abstained in 14 of 24 runs, and in the LangGraph program it reached
the turn limit without an answer in 19 of 24. GPT-5.6 Luna abstained
in 7 of 24 runs in each program.

\section{Discussion}
\label{sec:discussion}

The results separate two abilities. The first is to compute over the
data. With code, the strongest models answer nearly every computable
question of Section~\ref{sec:negative}. The second is to know which
records the company relies on and which rule applies to them. The
eight questions of this paper test the second ability, and the
strongest models pass only part of the test.

The three questions about several records show where the line falls.
In the conflicting-amounts question, one property on the HubSpot
record names the system of record, and Claude Fable 5.1 found it in
every run. In the roadmap and signed-proposal questions, nothing names
the right record. The agent has to notice that one project stopped
changing on a given day, or that only one proposal's price hold
covered the signing date. All twelve agents together answered these
two questions once in 72 runs. On the same two questions, the four
strongest agents made fewer tool calls than on any other question and
never abstained. The pattern is consistent with an agent that reads
the record the question names, finds a plausible value, and stops.

Code did not help on these questions, although it helped a great deal
on the computable ones. A sandbox makes it easy to read many records.
It does not tell the agent which of them the company relies on.

The missing knowledge is small. It is a handful of facts of the kind
that an employee learns in the first weeks: which system holds the
committed amount, on which day the roadmap moved, how the pipeline
review is computed. A written store of such facts would turn the
hidden facts of this paper into recorded ones. The benchmark can
measure how much such a store helps, by running the same agents with
and without it.

\section{Limitations}
\label{sec:limitations}

All results come from one company and one seed. The templates produce
questions for other seeds as well, and the eight questions of the
seed-43 company pass the checks of Section~\ref{sec:build}, but we
have not run agents on them. Each agent has 24 runs, so the 95\%
intervals are up to 37 percentage points wide, and most differences
between agents are not statistically supported. We have not measured
how well a person answers the eight questions from the systems alone.
The remark on each call is the same sentence in every company, so a
model could learn to recognise it.

The generated data has defects that a reader of the questions should
know. In the four templates with a remark, the remark is the last turn
of a call whose other turns do not refer to it. In the lost-deal and
dropped-expansion templates, the rule in the remark selects a record
by its assignee, priority and dates, not by its content.

The six models come from three providers, and the results may differ
for other models. The result files record, for each run, whether the
answer was correct, which of its values were correct, whether it was
an abstention, and how many tool calls were made. They do not record
the answers or the tool calls themselves, so we cannot say which wrong
answers the agents gave. The results describe the models and agent
programs available in September 2026.

\section{Conclusion}

Agents that can run code now answer most questions whose answers are
in a company's records. We extended the Era by Eon benchmark with
eight question templates whose answers depend on hidden knowledge: a
promise made on a call, the record that the company relies on among
several, or the rules behind an old report. For every generated
company, the benchmark fills each template from the company's own data
and computes the answer without a language model. The evidence for
each answer is in the systems, but not in the record that the question
names.

On the company of this paper, the best agent answered 18 of 24 runs,
and four of six models answered at most 6. Every agent found the rule
when a person stated it on a call, at least some of the time. Almost
no agent found the right record when only dates and small marks in the
data pointed to it: two such questions were answered once in 72 runs.
Code, which decides the computable questions, did not help here. The
next step is to give agents a written store of the facts that
employees know, and to measure with this benchmark how much of the gap
it closes.

\bibliographystyle{ACM-Reference-Format}
\bibliography{hidden_knowledge}

\end{document}